
\input phyzzx
\FRONTPAGE
\PHYSREV
\def\NU{\address {$^a$ Department of Physics, Northeastern University, Boston
Massachusetts 02115}}
\def\ifunam{\address{
$^b$ Instituto de F\'\i sica, Universidad
Nacional Aut\'onoma de M\'exico, Apdo. Postal 20-364, 01000 M\'exico D.F.}}
\def\utrecht{\andaddress {$^c$
Instituut voor Theoretische Fysica, Universiteit Utrecht,
Princentonplein 5, Postbus 80.006, 3508 TA Utrecht, the Netherlands}}
\def\abstract{\vskip\frontpageskip\centerline{\twelverm ABSTRACT}
              \vskip\headskip }
\def\author#1{\vskip 0.3cm \titlestyle{\twelvecp #1}\nobreak}

\def\sqr#1#2{{\vcenter{\vbox{\hrule height.#2pt
        \hbox{\vrule width.#2pt height#1pt \kern#1pt
          \vrule width.#2pt}
       \hrule height.#2pt}}}}
\def\square{\mathchoice\sqr34\sqr34\sqr{2.1}3\sqr{1.5}3}
 \nopubblock
\title { \bf CRITICAL EXPONENTS OF THE FULLY
FRUSTRATED 2-D XY MODEL}
\author {G. Ramirez-Santiago$^{a,b}$ and Jorge ~V. ~Jos\'e$^{a,c}$}
\NU
\ifunam
\utrecht
\vskip 0.3truecm
\centerline {ABSTRACT}
\par
We present a detailed study of the critical properties of the
2-D XY model with maximal frustration in a square lattice.
We use extensive Monte Carlo simulations to study the thermodynamics
of the spin and chiral degrees of freedom, concentrating on their
correlation functions.
The gauge invariant spin-spin correlation functions are calculated
close to the critical point for lattice
sizes up to $240\times 240$; the chiral correlation functions are studied
on lattices up to  $96\times 96$.   We find that the critical exponents
of the spin phase transition are  $\nu=0.3069$, and $\eta=0.1915$, which  are
to be compared with the unfrustrated XY model exponents  $\nu=1/2$
and $\eta=0.25$.
We also find that the critical exponents of the chiral transition
are $\nu_{\chi}=0.875$, $2\beta=0.1936$,
$2\gamma=1.82$, and $2\gamma\>\prime=1.025$, which are
different from the expected 2-D Ising critical exponents.
The spin-phase transition occurs at $T_{U(1)}=0.446$ which is
about 7\% above the  estimated chiral critical temperature $T_{Z_{2}}=0.4206$.
However, because of the size of the statistical errors,
it is difficult to decide with certainty
whether the transitions occur at the same
or at slightly different temperatures. Finally, the jump in the helicity
modulus in the fully frustrated system is found to be about 23\% below the
unfrustrated universal value. The most important consequence of these
results is that the fully frustrated XY model
appears to be in a novel universality class. Recent successful comparisons
of some of these results with experimental data are also briefly discussed.
\endpage

\chapter{Introduction.}
\vskip 0.4truecm
The critical behavior of the uniformly frustrated 2-D XY
model has been studied extensively in recent years, both theoretically$^{1-25}$
and experimentally$^{26-32}$. This theoretical interest has been due to
the rich variety of possible novel critical phenomena
that can appear in this model depending on the frustration parameter
$f=p/q$, with $p$ and $q$ relative primes.
Experimentally, an understanding of the phase transition(s) that occurs in this
model is important to describe the physics of
two-dimensional periodic arrays of Josephson junctions,$^{26-32}$ and
two-dimensional superconducting wire networks$^{33-34}$, both in the
presence of frustration $f=\Phi/\Phi_0$. Here  $\Phi$ is the average
flux per plaquette
normalized to the superconducting quantum of flux $\Phi_0 =h/{2e}$.
These arrays  can be manufactured with high precision using
modern photolithographic techniques. Of particular interest
is the $f=1/2$ fully frustrated
2-D XY model (FFXYM). This model  has a continuous U(1) abelian
symmetry, and a
discrete $Z_2$ symmetry leading to the possibility of true long-range
order in two dimensions. In contrast, the unfrustrated
2-D XY model (XYM) only possesses a
continuous U(1) abelian symmetry: its low temperature phase is
characterized by quasi-long range order rather than true long-range
order$^{35-38}$.  In spite of the many experimental and theoretical
studies of the FFXYM, there are several questions that remain to be
resolved. For example, it is not clear whether one
phase transition exists at the critical temperature $T_c$  which is
a combination of a Berezinskii-Kosterlitz-Thouless (BKT) type transition
for the U(1) symmetry plus an Ising-like transition for the $Z_2$ symmetry,
or whether there are two successive phase transitions at critical temperatures
$T_{U(1)}$ and $T_{Z_{2}}$. Even the order in which they may occur is
controversial. More importantly the nature of the transitions, as characterized
by their critical properties, is not yet fully understood.

\par
In their original work Teitel and Jayaprakash suggested$\>^{4}$
that in a square lattice the two transitions occurred very close in
temperature.
They carried out Monte Carlo (MC) simulations  to calculate
the helicity modulus $\Upsilon$ and the
specific heat $C$ as a function of temperature
and lattice sizes $L\times L$, with $L$ up to $L=32$.
They found that the maximum of the specific heat
appeared to increase as $\ln L$, characteristic of a 2-D Ising-like
transition. Related studies in the triangular lattice antiferromagnetic XYM,
which could be expected to be in the same
universality class as the FFXYM, indicate that there is a
combination of  BKT and Ising-like transitions$^{6}$.
In ref. ${6}$, the two transitions appear to take place at the same
$T_c$ while in ref.${7}$ they are within $2\%$ of each other, with
$T_{Z_2}>T_{U(1)}$.
In another investigation Berge {\it et al.}$^{14}$ introduced a frustrated XYM
with variable frustration
on a square lattice. In this model the couplings along the columns
are chosen with strength
$J$, while those along every other row have strength $- \mu J$,
with  $0< \mu \leq 1$. From a MC analysis of the specific heat they surmised
that for $\mu< 1$
the model has separate Ising and BKT ordering with
$T_{Z_2}<T_{U(1)}$, while for the FFXYM ($\mu=1$) the two transitions
appear to merge into one. The Berge et al. model  was studied in detail by
Eikmans et al.$^{16}$ who carried out MC calculations of the helicity
modulus, which is more sensitive to possible BKT-like ordering, and
interpreted their results using a Coulomb gas picture.
In another thermodynamic MC study of
the related  uniformly frustrated square lattice 2-D Coulomb gas model,
Grest$^{17}$ carried out simulations
for frustrations $f=0, \> 1/2,\> 1/3$ and $1/4$ and with $L$ up to 50.
He found that in the fully frustrated case the jump in the inverse
dielectric constant $\epsilon^{-1}_{o}$ is different from the XYM case.
Specifically, the jump in $\epsilon^{-1}_{o}$
occurs at $T_{CG}=0.129\pm 0.002$ and
takes the value $\epsilon^{-1}_{o}=0.63\pm 0.03$, which is larger than the
XYM universal value of $0.52$, and agrees with Minnhagen's conjecture.$^{39}$
The determination of the jump was based, however, on the criterion
used for the XYM. Grest found, as in previous studies, that the
specific heat  grows logarithmically with $L$. It is significant that
his results appear to indicate a clear separation of the two critical
temperatures with $T_{Z_2}>T_{U(1)}$, contrary to previous conjectures.

\par
Granato$\>^{19}$ {\it et al.} have studied  the $Z_2$
critical behavior of a coupled
XY-Ising system using MC and MC-transfer matrix calculations.
An important finding in this study is the chiral critical exponent
$\nu _{\chi}\sim 0.85(3)$, which is clearly different from the 2-D Ising model
value of $\nu=1$.$^{18}$ Furthermore, they found
that the XY and Ising transitions occur at essentially the same temperature.
Lee {\it et al.}$\>^{20}$ carried out MC simulations of the FFXYM
in the square and triangular lattices and found that $\nu$ is also different
from the 2-D Ising model result.

\par
In the XYM the nature of the BKT phase is characterized by the approximate
analytic expression for the spin-spin  correlation functions$^{36,37}$.
However,
unlike in the XYM case, it has proven to be very difficult to
calculate the correlation functions for the FFYXM analytically. This
difficulty exists partly because in order to carry out the
calculations one needs
to include the basic excitations of the frustrated problem, which
consist of different types of fractional charges as well as the
Ising model related domain walls$^{11}$.
Nonetheless, it has been possible to extract some qualitative
information  about the critical properties using techniques such
as the renormalization group approximation$\>^{9}$ applied to an effective
hamiltonian obtained from a Hubbard-Stratonovich
transformation of the FFXYM$^{8}$
or by general symmetry arguments$^{10}$. One worry about the effective
hamiltonian is that it does not explicitly contain the
same elementary excitations as the original FFXYM, such as the
fractional charges.
All of the studies mentioned above have mostly concentrated on calculating
thermodynamic quantities, for it has been difficult to separate the
$Z_2$ from the $U(1)$ contributions.

The purpose of this paper is to fill this gap by explicitly calculating
the $U(1)$ and $Z_2$ correlation functions as well as the separate $Z_2$
contribution to the magnetic properties. We should mention at the
outset that these calculations are significantly more demanding
than the thermodynamic calculations and are now possible because of
improved algorithms and  computer power. One further complication
is that at present there is no available analytic  theory for
the $f\not=0$ case that could suggest what form
these correlation functions should have and we need to make {an \it ansatz} for
them. Generally, we
can either assume that they decay exponentially or algebraically with distance.
We use different statistical measures to test for the two possibilities.
If our MC  results for the correlation functions are consistent with
an exponential  decay we extract
a correlation length $\xi (T)$, while if they are consistent with a power law
decay we extract the corresponding $\eta (T)$ exponent.
In the case that $\xi (T)$ diverges at $T_c$ from above it can diverge
as a power law or with the BKT form $\sim\>\exp\left(B(T-T_{c})^{-\nu}
\right)$.  In the $f=0$ case the critical exponent $\nu (f=0)=1/2$.$^{36,37}$
In the low temperature phase of the XYM the correlation function decays
algebraically with distance $r$ as $\sim r^{-\eta}$, where the  $\eta$
exponent
is  a continuous function
of $T$ and takes  the universal  value $\eta(f=0,T_{BKT})=1/4$.
Several experiments have confirmed the $f=0$ picture and
the values of the measured critical exponents$^{40-42}$ agree well with those
predicted by theory. In addition, recent MC simulations have
provided an accurate evaluation of the $f=0$ XYM critical exponents$^{43-47}$.
The most recent$^{46}$ high statistics estimates for $f=0$ are
$\> \nu=0.4695(1)$ and $\eta =0.235$, with the critical temperature
$T_{BKT}=0.8953$.
\par

In order to understand the nature of the phase transitions
in the FFXYM we have studied a variety
of quantities, several of which separately describe each particular symmetry.
The thermodynamic quantities calculated are:
the helicity modulus, $\Upsilon$, and the square of both the
staggered chiral magnetization, ${\cal M}^{2}_{s}$, and
susceptibility, $\chi^{2}_{s}$. We have carried out an extensive analysis
of the gauge invariant $U(1)$ correlation function, $g_{(U(1)}(r)$
 and their corresponding even and odd coherence lengths (to be defined below).
These calculations have allowed us to extract the $U(1)$  critical temperature,
$T_{U(1)}$, and its critical exponents $\nu$ and $\eta$. For the $Z_2$ freedoms
we calculated the chiral correlation function, $g_{\chi}(r)$, and its
corresponding coherence  length, $\xi_{\chi}$, which allowed us to estimate
the critical exponent $\nu_{\chi}$ and the
critical temperature $T_{Z_{2}}$.  Our result for the exponent
$\nu_{\chi}$ is in
very good agreement with the recent MC transfer matrix calculation$^{25}$.

We will now outline the main results of our study.
Our extensive analysis is consistent with a $U(1)$ BKT-type
transition but with exponents $\nu $(f=1/2)$=0.3069$ and
$\eta$ (f=1/2, $T_c)=
0.1915$. These results clearly differ from those obtained in the XYM case
$^{36,37, 43}$.
We have also calculated the $Z_2$ critical exponent  $2\beta=0.1936(35)$
for ${\cal M}^{2}_{s}$,
$2\gamma=1.82(13)$ and $2\gamma \prime=1.025(79)$ for $\chi^{2}_{s}$,
and the coherence length
exponent $\nu_{\chi}=0.875$. These exponents
are also different from  those expected for a 2-D Ising model.
The critical temperatures found in our study are $T_{U(1)}=0.446$
and $T_{Z_{2}}=0.4206$. One could be tempted to say that the transitions
take place at two different temperatures, and this may indeed be the case.
However, after a detailed assessment of the size of the statistical
errors from the nonlinear fits and considering the small
difference between the two temperatures we can not be certain if they are
different or not. Furthermore, the transitions are reversed
from their expected order. We suspect that more extensive simulations
with algorithm improvements, better statistics
and larger system sizes are needed to clarify this point.
The results mentioned above were obtained from extensive
MC simulations on a square lattice of size L, with L ranging from
$L=8$ up to $L=240$, and with periodic boundary conditions.
What emerges from our results is that the FFXYM is in
a novel universality class different from either a pure XY or an Ising
universality class. A brief description of some of the results presented
here has appeared elsewhere$^{22}$.

\par
We have also recently reanalyzed the experimental results of the Delft
group$^{48}$ for $f=0$ and $f=1/2$. We have concluded that the values of
$\eta (f=0)=1/4$ and $\eta (f=1/2)= 0.1915\approx 1/5$  are in good agreement
with the experimental data. However, the fits of the experimental
resistance versus
temperature data can not distinguish between a  $\nu (f=1/2)=1/3$
from a $\nu (f=1/2)=1/2$. Moreover, as mentioned above,
recent MC-transfer matrix work has provided further evidence that
the chiral exponents  are not equal to the 2-D Ising model exponents and even
quantitatively the value of $\nu_{\chi}$ has begun to converge on
values close to $0.85$.

\par
The organization of this paper is as follows:
In section 2.1 we define and briefly review the general
properties of the uniformly frustrated 2-D XYM. In section 2.2
we define the thermodynamic quantities calculated in this paper
while in section 2.3 we give the expressions for the calculated gauge invariant
$U(1)$ and $Z_2$ zero-momentum correlation functions, of central interest
here, together with their possible asymptotic behaviors.
In Section 3.1 we describe briefly the MC algorithm used in our calculations.
Since there are no analytic results for the correlation functions to guide
our analysis, we proceed by developing an approach that consists of
using several independent checks  of  the results obtained. As a test,
in subsection 3.2, we successfully apply our
strategy  to the unfrustrated XYM and compare our results to those
obtained in the more
extensive recent  MC studies$^{43-46}$.
In section 4 we present the bulk of our numerical MC results
applied to the $FFXYM$. In 4.1 we discuss the thermodynamic results
for both the $U(1)$ and $Z_2$  freedoms. In 4.2(a) we give the correlation
function results for the $U(1)$ freedoms, including a finite size
scaling analysis for the correlation length. In section 4.3(b) we present the
corresponding correlation function results for the $Z_2$ freedoms.
Finally in section 5 we present a critique of our results and a possible
outlook for the future.

\chapter{\bf The fully frustrated XY model}
\smallskip
\section{\bf Definition of the model.}
\vskip 0.3truecm
The uniformly frustrated 2-D XYM is defined by the hamiltonian
$$H=-\sum_{<\vec r, \vec r \prime >} J
\cos\Bigl( \theta (\vec r)- \theta (\vec r \prime ) +
f(\vec r, \vec r \prime)\Bigr),
\eqno(1)$$
where $\theta (\vec r )$ is the angle at site $\vec r$,
$<\vec r, \vec r \>\prime>$ stands for a sum over nearest-neighbor lattice
sites, and $J$ is the exchange constant.
In the  Josephson junction array  representation of  the model
in a transverse magnetic field, the bond variables $f(\vec r ,
\vec r \prime )$ are given by the line integral
$f(\vec r , \vec r \> \prime)={{2 \pi }\over { \Phi_{0} }}
\int^{\vec r\> \prime}_{\vec r}
\vec A \cdot d\vec l $,
with $\vec A$ the magnetic vector potential. For uniform
frustration these bond angles are required to satisfy
$$\sum_{plaquette} f(\vec r, \vec r \> \prime)={{2\pi}\over \Phi_{o}}
\oint_{plaquette} \vec A \cdot d\vec l = 2\pi f. \eqno(2)$$
The hamiltonian defined in Eq. (1) is invariant under
the transformation,
$\theta (\vec r)\rightarrow \theta (\vec r)+2\pi n(\vec r)$ and
$f(\vec r, \vec r \> \prime )\rightarrow f(\vec r, \vec r \> \prime)+
2\pi [n(\vec r\prime ) - n(\vec r )]$,
where $n(\vec r)$ and $n(\vec r \prime )$ are integer numbers.\break
Choosing the gauge $\vec A=(-By,0,0)$ so that $\vec B=B {\bf \hat z}$
and assuming a square
lattice, the bond angles $f(\vec r, \vec r \prime)$ are given by
$$\eqalignno{ f(\vec r, \vec r\>\prime) &= \mp2\pi f(j+{{1}\over{2}}),
\>\>\> \hbox{for}\>\>\> \vec r'=\vec r \pm a_{0} \vec i
\>\>\>\>\>\hbox{and}  \>\>\>\>\> & (3a) \cr
f(\vec r, \vec r\>\prime) &= 0\>\>\> \hbox{for}
\>\>\>\vec r'=\vec r \pm a_{0} \vec j.
& (3b) \cr }$$
Here $f={{Ba_{o}^{2}}/{\Phi_{o}}}\>$ with  $\vec r =(ia_{o},ja_{o})$,
$i,j$ integers and  $a_{o}$  the lattice spacing.
The uniformly frustrated model is periodic in $f$ with period one, and
with reflection symmetry about $f=1/2$. The XYM corresponds to the
unfrustrated $f=0$ case.
\par

The fully frustrated case corresponds to $f=1/2$.
The effect of $f$ in this case is to
produce alternate rows with ferro- and antiferromagnetic
couplings, while the couplings along the columns are all ferromagnetic.
Each plaquette has  one antiferromagnetic and three ferromagnetic
bonds, or vice versa, leading to a ground state that has a  two-fold degeneracy
with half-integer vortices of opposite circulation or chirality.$\>^{1}$
Thus, the system displays two symmetries: the underlying continuous
$U(1)$ abelian symmetry for the phases and a discrete
$Z_2$ or Ising-like symmetry associated with  the chiral degrees of freedom.

\bigskip
\section{\bf Thermodynamic properties}
\par
The helicity modulus $\Upsilon$ is defined by the
response of the system to a twist in the spins at its boundaries.
In our case $\Upsilon$ is calculated explicitly from
the formula,
$$\eqalignno{\Upsilon&={{1}\over {N}} \Biggl[ \VEV{
\sum_{<\vec r,\vec r \> \prime >} x^{2}_{\vec r, \vec r \> \prime }
\cos \Bigl( \theta (\vec r)- \theta (\vec r \> \prime )
+f(\vec r, \vec r \> \prime) \Bigr)} \cr
&-{{1}\over {k_BT}}\VEV{ \Bigl[\sum_{<\vec r, \vec r \> \prime>}
x_{\vec r, \vec r \> \prime} \sin \Bigl( \theta (\vec r )-
\theta (\vec r \> \prime )+
f (\vec r, \vec r \> \prime ) \Bigr) \Bigr] ^{2} } &(4) \cr
&+{{1}\over {k_BT}}\Bigl[ \VEV{ \sum_{<\vec r, \vec r \> \prime >}
x_{\vec r, \vec r \> \prime} \sin \Bigl( \theta (\vec r)-
\theta (\vec r \> \prime)+
f(\vec r, \vec r \> \prime ) \Bigr)} ^{2} \Bigr] \Biggr], \cr
}$$
where $\VEV {\hbox{     }}$ stands for a thermal average, and $k_B$ is
Boltzmann's constant, and
 $x_{\vec r, \vec r'}=x_{\vec r}-x_{\vec r'}$. Another  informative
quantity associated with the
supercurrent loops around a plaquette in a Josephson array is the
staggered magnetization
$$M_{stagg.}={{1}\over {N}} \VEV{ \sum_{{\cal P}(\vec R)}
(-1)^{R_{x}+R_{y}} \Bigl[
\sum_{<\vec r, \vec r \> \prime > \in {{\cal P}(\vec R)}}
\sin \Bigl( \theta (\vec r )-\theta (\vec r \> \prime )
+f(\vec r, \vec r \> \prime) \Bigr) \Bigr]
}, \eqno(5)$$
where $(R_x, R_y)$ give the coordinates at the center of the plaquette
${\cal P}$. The index ${\cal P}(\vec R)$ runs from
1 up to $N$, the total number of plaquettes in the lattice.
\par
We now turn to the definition of the
quantities associated with the chiral degrees of freedom.  The {\it chirality}
of a plaquette gives the direction of circulation of the supercurrents induced
by frustration. Each plaquette has a definite {\it chirality}
which can be $\pm1$, and it is calculated from$^{1}$
$$\chi (\vec R)= \hbox{sign} \Bigl[ \sum_{<\vec r, \vec r \> \prime>\in
{\cal P}(\vec R)}
\sin \Bigl( \theta (\vec r ) - \theta (\vec r \> \prime)+
f(\vec r, \vec r \> \prime) \Bigr) \Bigr], \eqno(6)$$
with the dual lattice vector $\vec R=[(i+1/2)a_{o},(j+1/2)a_{o}]$ with
$i,j$ integer numbers. At zero temperature
the chiralities are ordered like
a 2-D Ising antiferromagnet. At finite temperatures there are line or
domain wall defects separating regions with different chiralities.
\par
The order parameter describing the $Z_{2}$ phase transition is the
{\it staggered chiral magnetization}, defined by
$${\cal M}_{s}=\VEV { {{1}\over {N}} \sum_{\vec R} (-1)^{R_{x}+R_{y}}
\chi (\vec R)} \eqno(7)$$
It is difficult to study this quantity numerically
since it oscillates rapidly between positive and
negative values. A more stable quantity to study is
 Binder's second order cumulant, ${\cal M}^{2}_{s}$
and its fluctuations.$^{7}$ These fluctuations are given by the square
of the {\it staggered chiral susceptibility }, calculated as
$$\chi^{2}_{s}=N\VEV{ { { { \Delta {\cal M}^{2}_{s} } }
\over {k_{B}T} } }. \eqno(8)$$
\par
Our numerical results for the thermodynamic quantities
characterizing the chiral degrees of freedom will be presented in
section 4.1(b). It will be
seen that close to the critical region their behavior is quantitatively
different from an Ising ferromagnet on a square lattice.
\medskip
\section{\bf Correlation functions}

It is known that the hallmarks of the BKT ordering can be given in terms of
the phase correlation functions. An
analytic evaluation of these quantities appears to be mathematically
intractable for uniform frustration. Nonetheless, for random frustration it
has been possible to analytically calculate the
correlation functions at low temperatures in the limits where the density
$x_{f}$ of frustrated plaquettes$^{3}$ is $x_{f}\ll1$ or
$x_{f}\sim {1\over 2}$.
Given that the hamiltonian of the frustrated XYM is gauge invariant,
the phase correlation functions should also be gauge invariant.
The correlation functions  are defined
along a path connecting the correlated spins and are therefore path-dependent.
In fact, gauge invariant correlation functions along two
different paths differ by the total amount of frustration enclosed by the
two paths. The  gauge invariant phase correlation function along a path
$\Gamma$ that
accounts for the frustration in the system is given by$^{2,3}$
$$g_{U(1)}(\vec r, \vec r \> \prime)= \VEV{ e^{i \theta(\vec r )}
\Bigl( \prod_{<\vec s, \vec s\> \prime > \in \Gamma}
e^{i f(\vec s, \vec s\> \prime)} \Bigr)
e^{-i \theta(\vec r \> \prime)} }. \eqno(9)$$
In Fig. 1 we show two possible trajectories $\Gamma _{A}
(\vec r, \vec r \> \prime)$ and $\Gamma _{B} (\vec r, \vec r \> \prime)$
joining the points $\vec r$ and $\vec r \> \prime$. The phase
introduced in Eq.(9) when going around the trajectory A is
$\prod_{\Gamma_{A}} e^{i f(\vec s, \vec s \> \prime)}=
\exp \Bigl( \sum_{\Gamma_{A}} f(\vec s, \vec s \> \prime ) \Bigr)$,
while if one follows trajectory B the phase factor is given by\break
$\prod_{\Gamma_{B}} e^{i f(\vec s, \vec s \> \prime)}=
\exp \Bigl( \sum_{\Gamma_{B}} f(\vec s, \vec s\> \prime) \Bigr).$
The total frustration contained in the area enclosed by both trajectories is
then
$$\sum_{<\vec s, \vec s \> \prime> \in \Gamma_{A}} f(\vec s, \vec s \> \prime)
+ \sum_{<\vec s, \vec s \> \prime> \in \Gamma_{B}} f(\vec s, \vec s \> \prime)
=2\pi Mf, \eqno(10)$$
where $M$ represents the number of elementary plaquettes inside the
area encircled by the paths, and $f$ is the frustration of each
plaquette.  In this case the phase in Eq.(9) is shifted
by an additional amount $2\pi Mf$ when the correlation along trajectory B is
calculated instead of A.
\par
To evaluate the effect of frustration on the correlation
functions at low temperatures one can use duality  transformations to obtain
a lattice Coulomb gas representation of the model.
After doing so one obtains$^{3}$
$$g_{U(1)}(\vec r, \vec r \> \prime)= \exp \Bigl[
i \sum _{\vec r} \sum_{\vec R} {1\over 2} n(\vec r) \Theta (\vec r- \vec R)
f(\vec R)  \Bigr]  \> \times \>
g_{c.g.}(\vec r, \vec r \> \prime), \eqno(11)$$
where $g_{c.g.}(\vec r, \vec r \> \prime)$ is the lattice Coulomb gas
correlation function for the XYM$\>^{37}$ and $f(\vec R)$
is the frustration at the plaquette with center at the dual lattice site
$\vec R$. The number $n(\vec r)$ is zero everywhere, except at $\vec r$
and $\vec r \> \prime$, where it takes the values $n(\vec r)=
-n(\vec r \> \prime)=1$.$^{3}$
The angular potential $\Theta (\vec R)$ is given by$^{37}$
$ i \Theta (z)= \ln (z) - G(\vert z \vert),$
for large $R$. Here $z=R_{x}+i R_{y}$ and $\vec R=(R_{x}, R_{y})$.
Notice that in Eq.(11) there is an extra phase factor
appearing in the phase correlation function.
This factor weights the contributions to the correlation function
coming from different trajectories going between $r$ and $r'$,
and it is a consequence of gauge invariance
or, equivalently, the Aharanov-Bohm effect. We shall see that this
extra phase factor in the correlation function appears naturally
in the results discussed in Section 4.1(a).
\par
In the evaluation of the correlation functions
we can have important contributions from more than
one Lyapunov exponent, which makes the extraction of the largest exponent
difficult. However, this problem is not present if we evaluate the zero
momentum
correlation function defined by$^{43}$
$$g_{o}(r)=<\vec S_{av}(i) \cdot \vec S_{av}(i+r)>,\eqno(12)$$
where
$$\vec S_{av}(i)={1\over L_y} \sum_{j} \vec S(i,j)\eqno(13)$$
is the average spin along the $i$th-column.
In this case $r$ denotes the distance between the columns
being correlated.
\par
{}From the definition of the gauge-invariant phase correlation function,
Eq. (9), the expression for the zero momentum correlation function is
$$g_{U(1)}(r)=\VEV{ {{1}\over {L_{x}L_{y}}} \sum_{j=1}^{L_{y}}
\sum_{i=1}^{L_{x}}\cos \Bigl( \theta_{i+r,j}-\theta_{i,j}+
\pi (j+{1\over 2})r \Bigr) },\eqno(14)$$
where we have used Eqs.(3).
\par
A similar reasoning applies to the zero momentum
chiral correlation function given by
$$g_{{\chi}}(r)=\VEV {
{{1} \over {L_{x}L_{y}}} \sum_{i=1}^{L_{x}-1}
\sum_{j=1}^{L_{y}-1} \chi_{i+r,j} \> \chi_{i,j}}. \eqno(15)$$
We have used equations (14) and (15) to evaluate the correlation
functions in our numerical simulation.

\par
As mentioned before there are no  known
explicit analytic expressions for $g_{U(1)}(r)$ and $g_{\chi}(r)$.
Nonetheless, we can make an {\it ansatz} for the analytic form
of these correlations close to the critical region. Our {\it ansatz}
is based on what is known about the XYM and about general properties of
standard second order phase transitions. In
the XYM as $T \rightarrow T^{+}_{BKT}$ the asymptotic form
of the spin-spin correlation function for $r\gg 1$ is
 $$g_0(r)=
{{{\cal A}_0}\over {r^{\eta_0}}}\> \times \> e^{-{{r}
\over {\xi_0}}},\eqno(16)$$
with  $\eta_0 =1/4$ and the coherence length diverging as
$$\xi_0(T)=A_0\> \exp \Biggl( {{B_0}\over {(T-T_{BKT})^{\nu_0}}} \Biggr).
\eqno(17)$$
In the XYM the critical exponent $\nu_0=1/2$.$^{36}$
In the low temperature phase ($T\leq T_{BKT}$)
the long distance correlation function decays as
$$g_0(r)={{C_0} \over {r^{\eta _0(T)}}}.\eqno(18)$$
The exponent $\eta_0$ is a function of temperature, representing
a continuous line of critical points. In the XYM $\eta_0$ takes the universal
value $\eta_0 (T=T_{BKT})=1/4$.$^{36,37}$
This result is directly related to the universal jump predicted for the
superfluid density.$^{38}$
\par
 In the disordered phase  of the 2-D Ising model
($T>T_{I}$), the asymptotic behavior of the correlation function close to
$T_{I}$ is given by$^{49}$
$$g_{I}(r)=
{{A_{I}}\over {r^{\eta_{I}}}} \> \times \> e^{-{r\over \xi_{I}}}
\eqno(19)$$
again with $\eta _{I}=1/4$ and with
a power law divergence for the coherence length
$$\xi_{I}(T)={{A_{I}}\over {(T-T_{I})^{\nu_{I}} } },\eqno(20)$$
where the critical exponent $\nu_{I}=1$.
In the ordered phase ($T \leq T_{I}$), the asymptotic  behavior of
the correlation functions is given by$^{49}$
$$g_{I}(r)={{C_{I}}\over {r^{\eta_{I} } }}\times
e^{-{r\over \xi_{I}}}+<M_{I}>^{2}, \eqno(21)$$
valid for $\epsilon_I={{T_{I}-T}\over {T_{I}}}<<1$ and $\epsilon_I \>r <1 $.
The correlation function critical exponent at $T_{I}$ is $\eta_{I}=1/4$,
as in the XYM.
\par
Based on previous studies of the thermodynamics of the FFXYM
it is reasonable to assume that their asymptotic
behavior for $f=1/2$ can be described by either BKT or Ising-like
forms described above. In our calculations
we checked for the best fits to our MC data by either form.
\par
Having discussed the expected analytic forms for the different correlation
functions of interest, let us now turn to a discussion of their numerical
evaluation. We should notice first that, strictly
speaking, for finite lattices the asymptotic behavior of the correlation
functions is not accessible. Even in rather large lattices
the subleading power law behavior of the correlation functions can be
non-negligible. Thus, the evaluation of the
coherence length extracted from a numerical calculation of the correlation
function is nontrivial. A common procedure is to take periodic boundary
conditions and then fit the behavior of the correlations to
$$G(r)=g(r)+g(L-r),\eqno(22)$$
where $g(r)$ is any of the correlation functions of interest.
We should note that in general it is
not sufficient to account for the closest images to the source along
the $r$ direction but we must also account for the images farther away
as well as those in the  direction transverse to $r$. In
fact, their contribution becomes more important as the coherence length
$\xi$ grows  since the number of relevant images increases.

\par
\chapter{ \bf Calculational strategy and test}
\vskip 0.5truecm

Our strategy is to carry out several independent consistency checks of
our results,
for there are no analytic results with which to guide the analysis.
To test the reliability of
our consistency checks, we start by applying them to the extensively studied
XYM. Although there is a basic consensus about the physical
nature of the BKT transition, relatively reliable and thorough
nonperturbative numerical studies of the critical exponents
of the XYM became available  just recently.$^{43-47}$
Here we tried to follow some of the basic ideas of these approaches, in
particular the one used above $T_c$, complemented with other tests
implemented here. We must stress that we are on less
firm ground in the FFXYM case than in the XYM and thus we need extra
consistency checks that were not needed in the XYM studies.
\vskip 0.5truecm
\section{\bf MC algorithm}
\vskip 0.5truecm
Different acceleration algorithms that have worked out well
in the XYM were constructed with special regard to
the nature of the basic excitations in the model. In the FFXYM we have a less
definitive idea about the basic excitations in the model
and therefore the same type of algorithms have not proven any more efficient
than the standard Metropolis approach.$^{49,50}$
Our simulations were then carried out using the standard Metropolis algorithm
in square lattices of sizes $L\times L$  with
$L=8, 16, 32, 60, 72, 84, 96, 180$, and $240$,
with periodic boundary conditions. The lattice sizes $L=120$, and $L= 180, 240$
were considered only at two temperatures that  are about 3\% and
2\% away from the estimated critical point, respectively.
These relatively large lattices were studied in order to get a better
estimate of the $U(1)$ correlation function  critical exponents. For
the thermodynamic and chiral degrees of freedom a full set of temperature
values was considered for $L$ up to $L=84$.
Since we expect to have critical slowing down similar to that seen in the XYM,
which has decorrelation times growing as $\xi^2$, we performed
reasonably long runs, although no detailed attempt was made to calculate the
FFXYM dynamic critical exponent. Nonetheless, our consistency and
self-consistency checks give support to the reliability of our results.
The equilibration time of a typical run was at least 10K MCS/angle far
from criticality and at least twice as much for temperatures
close to $T_c$. The statistics were calculated from runs
of at least 50K MCS/angle, and up to 290K MCS/angle.
Details of the length of the runs are given in the tables.
It is important to note here that since we are interested in extracting
critical exponents from nonlinear fits, plots of the results are sometimes
not as informative as looking at the numbers themselves.
\vskip 0.5truecm
\section{\bf Unfrustrated 2-D XY model}
\vskip 0.5truecm
We begin by presenting our results for the XYM together with the tests
of our consistency checks for both $\Upsilon ^0$ and the correlation functions.
We compare our XYM results to those obtained recently by more extensive
analysis$^{46}$. In the next section we shall present the bulk of the
results of our calculations with the FFXYM.
Note that in terms of the correlation
functions the difficult calculations are those for the $U(1)$ symmetry, for
in that case the correlation length may diverge exponentially rather than
algebraically as one gets close to $T_c$.

\par
We now describe the numerical approach to calculate the coherence
length, the critical temperature and critical exponents from the
$U(1)$ correlation functions. We
have to some extent repeated the recent, more extensive calculations
for the XYM$\>^{43-46}$ critical exponents for $T>T_{BKT}$, and  we
have extended the
calculations to the $T\leq T_{BKT}$ region. To facilitate the comparison
of our results, given in Table 1A, to previous findings
we have summarized the results of references 43-46 in Tables 1B and 1C.
Basically, we have followed the method of analysis used in those references,
although the lattices we have simulated are not as large as the ones
considered there. However,
to reduce the finite size effects and to insure meaningful results for the
correlation functions, we kept the ratio ${L/\xi_0} \geq 4$ in all the
calculations. We note that, even though  the critical temperatures in the
FFXYM is about $T_{BKT}/2$, we kept this ratio at ${L/\xi} \geq 5$.
The calculational procedure is the following: First, the periodic form
of the zero-momentum correlation function $g_0(r)$ given in
Eqs. (14) and (22), with $f=0$, was calculated in the high temperature
phase. Next, we carried out unconstrained 3-parameter
nonlinear fits of the data to the form given in Eq(16). From these fits
we determined $\xi _0(T)$ and the
parameters $\eta_0$ and $A_0$. To further check the consistency of the
nonlinear fits we performed linear fits to the MC data of
the form
$$\ln \biggr (g_0(r) \biggr)=\ln A_0
+\ln \biggr [ r^{-\eta_0}\>e^{-r/{\xi_0}}+
(L-r)^{-{\eta_0}}e^{-(L-r)/{\xi_0}}\biggr ]\eqno(23)$$
varying the values of $\eta_0$
until we reached a minium for the $\chi ^2$ function. We found that $\xi_0(T)$
is systematically above the values of Ref. 43 (for the comparison
see Fig.2), and that $\eta_0(T)$ oscillates nonmonotonically close
to the critical point, making its determination difficult.
We also carried out an unconstrained 4-parameter
nonlinear fit to the data to extract
$T_{BKT}$ and $\nu_0$, assuming a BKT form for $\xi _0(T)$.
As pointed out in a recent finite size scaling analysis$\>^{45}$
of the  XYM susceptibility, it is very difficult to distinguish
between a BKT form and a
power law divergence on the basis of data obtained from
MC simulations  on lattices up to L=256, even for temperatures 1\%
away from the critical point.
Thus, we carried out an additional unconstrained 3-parameter
nonlinear fit to a power law form [Eq. (20)].
The results of these fits are summarized in the first and
second columns of Table 1A. Before proceeding with a
discussion of these results,
let us emphasize that we carried out a careful analysis of the stability
of the parameters obtained from the fits by trying to make sure that the values
obtained correspond to the minimum of the $\chi^{2}$ function, within
the statistical errors of our simulations.
More details of the fitting procedure and related analyses
will be discussed below.

Let us now turn to the results obtained by assuming a BKT form for $\xi _0(T)$.
We find that the values of the parameters $A_0$ and $B_0$ in the first
column of Tables 1A and 1B agree well,
while the values of $\nu_0$ and $T_{BKT}$ differ by 4\% and 0.75\%,
respectively. The results obtained assuming a power law form
are given in the second column of Tables 1A and 1B.
The values of $T_{c}$ and $\nu_0'$ are within 1\% and 9\%, respectively,
whereas the values for the $A_0'$'s are the same. All of these results indicate
that there is good agreement between our results and those obtained
from the more  extensive MC simulations, in spite of the fact that our
simulations are in  smaller systems and have less statistics.
It is important to realize that fitting the coherence
length to a BKT or to a power law form leads to $\chi^{2}/dof$-values
of comparable quality. Thus, from this analysis alone
 one cannot decide which of the two fits is the correct one.
To sort out this problem we also have calculated the low temperature
$g_0(r,T)$.
The correlations,
for $L=60$ lattices, were fitted to $g_0(r)\sim  r^{-\eta_0}$ and
$$g_0(r)\sim \left[ r^{-\eta_0}\>e^{ \alpha_{0} r}+
(L-r)^{-\eta_0}\>e^{\alpha_{0} \left( L-r \right)} \right],\eqno(24)$$
at 9 different temperatures.
The results of these fits are given in Tables 2A and 2C.
It is found that the exponential fits appear to be  of better quality,
and the corresponding  values of $\eta_0(T)$ are systematically above
those obtained from the algebraic fits. However, the important point
is that the values of the exponents $\alpha$ are smaller
than 10$^{-2}$, suggesting a coherence length too large for the fit to be
trusted. Moreover, the algebraic fits agree with the low $T$
spin wave prediction, $\eta_0(T)\sim T$. In Fig.3(a) we show the results
for $\eta_0$ as a function of temperature.
To further analyze the nature of the low temperature
phase, we calculated  $\eta_0(T_{BKT})=\eta_{0c}$ as a function of lattice
size,
using the $T_{BKT}$ obtained at high temperatures and for
reasonably long runs. On the other hand, the exponential fits yielded values
of $\eta_{0c}$ systematically above those calculated from the
algebraic fit, shown in Fig.3(b). Fits to the exponential form plus
its image lead to better results in this case. We found that the values
of $\alpha$ are quite small and it appears that they become smaller as
L increases. This  suggests
that the algebraic contribution will dominate in the asymptotic limit.
Notice also that $\eta_{0}$ increases slowly with
$L$ and it does not seem to saturate for the larger lattices in both
types of fits.  The values for the exponents calculated from both types of fits
in the largest lattices  do agree with those obtained in
Refs.$\>44$ and ${46}$, with  the later results given in Table 1C for
comparison.  The value $\eta_{0}=0.2386(2)$ calculated from our algebraic fits
is in very good agreement with MC renormalization group
calculations.$\>^{46}$  However, the value $\eta_{0}=0.2713(2)$
obtained by assuming an exponential fit plus its images at
$T=T_{BKT}$ also agrees with the results$\>^{44}$ obtained studying the
relationship between $\xi_0(T)$ and the susceptibility $\chi(T)$
as $T\rightarrow T^{+}_{BKT}$.
The results from the present analysis
suggest, as expected, that the algebraic form gives a better fit
in the low temperature phase.

As a further check to this conclusion, to be used in the FFXYM analysis,
we now show that the values of $T_{BKT}$, $\Upsilon^{o}$
and $\eta_{0}$, which were determined independently,
{\it are consistent with
the universal value for the jump in} $\Upsilon^{o}(T=T_{BKT})$.
We calculated the magnitude of the jump in $\Upsilon^{o}(T=T_{BKT})$,
for  sizes $L=8,\> 16,\> 24,\> 32,\> 48$, $60,\> 72,\> 84$,
and $96$, and carried out a finite size analysis. We used
the $T_{BKT}=0.9035(6)$ obtained from the high temperature analysis
of the $U(1)$ correlation functions, to be described below.
The results are given in Table 3A.
We got the extrapolated value of
$${\Upsilon^{o}(T_{BKT})={0.5986(49)}}\eqno(25)$$
for the infinite system, by fitting a  straight line to
$\Upsilon^{o}(T_{BKT})$ vs
$L^{-1}$, and using the $L=48$ to $96$ results (See Fig.4(a)).
We also estimated the magnitude of the jump from the
universal intercept of $\Upsilon ^{o}(T_{BKT})$ with the ${2T_{BKT}}/\pi$ line
getting
$$\Upsilon^{0}(T_{BKT})=0.5752(4),\eqno(26)$$
 which is about 3\% below the value in Eq(25). Next, we considered
 the relationship between the exponent $\eta_0$ and the universal prediction
for the jump in $\Upsilon^{o}(T_{BKT})$$\>^{38}$, that is
$\eta_{0}=k_{B}T_{BKT}/[2\pi\Upsilon^{o}(T_{BKT})]$.
Before inserting the numbers it is important to stress the fact that the three
quantities appearing  in this  equation were  obtained from three
different calculations. The critical temperature was calculated from
the coherence length analysis in the high temperature phase,
$\Upsilon^{o}(T_{BKT})$ from a finite size analysis at $T_{BKT}$,
and the $\eta_0$ from a finite size analysis of
the algebraic correlation functions  at  $T_{BKT}$. Plugging in the numbers
gives the result

$$\eta_{0}=0.2402(18),\eqno(27)$$
 which  indicates that  the values of these quantities
satisfy, within the errors, the universality relation for the jump in
$\Upsilon^0$ at $T_{BKT}$.

In conclusion, we have shown in this section that our strategy yields
reasonable quantitative estimates of
the critical temperature, critical exponents and the magnitude of the jump
of $\Upsilon^{o}(T_{BKT})$ in the XYM. It is reassuring that
independent calculations lead to essentially the same quantitative results.
Building from what we have learned in this section about the XYM,
in the next section we proceed to apply the same logic and analysis to the
study of the phase transition(s) in the FFXYM.
\vskip 0.5truecm
\chapter{\bf Critical properties of the fully frustrated 2-D XY model}
\vskip 0.3truecm
In this section we present the bulk of our thermodynamic and correlation
function results for both the $U(1)$ and $Z_2$ freedoms. We start by
discussing the thermodynamic properties and then we move on to present our
results for the correlation functions.

\section{\bf Thermodynamic properties}
\vskip 0.3truecm
\subsection{\bf (a) U(1) freedoms}
\vskip 0.3truecm
We begin by discussing the helicity modulus $\Upsilon (T)$.
Previous studies$\>^{4,39}$ of the FFXYM
and the fully frustrated Coulomb gas$^{17}$ have indicated
the possibility that the
jump in $\Upsilon (T)$ may be different from the universal
XYM result. To further shed light onto this problem
we have studied $\Upsilon(T)$ as a function of
temperature for different lattice sizes and carried out a finite size
analysis of $\Upsilon_{U(1)}=\Upsilon(T=T_{U(1)})$.
Figure 5 shows the results in the temperature range $0.20<T<0.65$
obtained from runs for $L=8, \, 16, \, 32$ with 250K MCS and $L=60$ with
200K MCS. Notice that at low temperatures the finite size effects are almost
negligible, however, they become important in the critical region.
The behavior for L=32 and 60 is about the same in the temperature region
where $\Upsilon$ was calculated.
To investigate the magnitude of the jump in $\Upsilon$
we proceed as in the XYM calculations.
We performed a finite size analysis of $\Upsilon$ at
the  critical temperature $T_{\hbox{U(1)}}=0.44$, found from a high
temperature  analysis
of the correlations, to be discussed later. The simulations were
carried out in lattices of size $L=8,\, 16,\, 24,\, 32,\,
48,\, 60,\, 72,\, 84$, and $96$ and the
results are given in Table 3B. The behavior of
$\Upsilon_{U(1)}$  as a function of $1/L$ is shown in Fig. 4(b)
for $L=96,\, 84,\, 72,\, 60$ and $48$. The value
$$\Upsilon_{U(1)}=0.37(1)\eqno(28)$$
was estimated by extrapolating the data to an infinite lattice.
This  result suggests that for the lattice sizes and statistics of our
simulations, the jump in the helicity modulus for the FFXYM is about
 23\% below the XYM result.
The estimate $\Upsilon (T=T_{CG})=0.34(1)$  was obtained
from MC simulations of the fully frustrated {\it Coulomb gas} on a square
lattice$\>^{17}$ using the formula
$\Upsilon={ {T_{CG}  / {[2\pi \epsilon_{c}T_{CG} }}]}$
with the values $\epsilon_{c}^{-1}=0.63(3)$, and $T_{CG}=0.129(2)$.
Here, $\epsilon_{c}$ is the value of the dielectric constant at $T_{CG}$.
Thus, our extrapolated value for $\Upsilon_{c}$ and the one
obtained from the Coulomb gas data differ by about 7\%, which can be
considered in reasonable agreement.
Note, however, that one cannot rule out the possibility of a smaller value
of $\Upsilon_{T_{CG}}$ for larger lattices. Nonetheless,
we do not believe that the trend would change significantly from the result
given here. This result confirms previous
suggestions$^{4,17}$ and gives support to Minnhagen's heuristic
conjecture$^{39}$ about the difference between the jump of
$\Upsilon_{T_{CG}}$  for the frustrated Coulomb gas in a square lattice
as compared to the XYM universal jump.
\par
It has also been suggested$\>^{14}$  that the transition in the FFXYM could
be weakly first order. To check this possibility we looked at the
histogram of the energy about $T_{U(1)}$ and found no evidence for the
existence of two competing states.
In previous MC simulations$^{4,6,7,14,15,17}$ it was found that the
behavior of the maximum of the specific heat as a function of lattice
size was consistent with a logarithmic divergence, favoring an Ising-like
transition.  However, MC simulations in larger lattices$\>^{20}$
suggest that it is very difficult to
distinguish between a logarithmic or a power law divergence. We have studied
the specific heat  and found no signature for a logarithmic divergence
but we were unable to extract reliable exponents.
\par
We have also studied the staggered magnetization  $M_{stagg.}$
(Eq. (5)) due to the
supercurrents circulating around the plaquettes as a function of temperature
and lattice size. Figure 6 shows the behavior
of $M_{stagg.}$ as a function of $T$ for $L=16$ and $32$.
It is non-zero at low temperatures and drops sharply at about
$T=0.42$. We note that finite size effects are almost negligible for these
lattice sizes. The behavior of  $M_{stagg.}$ as a function of temperature
suggests that it may be considered as an order parameter for the $U(1)$
phase transition. Note, however, that the chirality is defined
in terms of the direction of the circulating  currents about the plaquettes
and thus $M_{stagg.}$ can also be thought of as an order parameter
for chirality.
\vskip 0.3truecm
\subsection{\bf (b) $Z_2$ freedoms}
\vskip 0.3truecm
We calculated the staggered chiral magnetization ${\cal M}_{s}$
defined in Eq(7). However, ${\cal M}_{s}$
oscillates too irregularly between positive and negative values, thus
it was more convenient instead to study$^{7}$
${\cal M}^{2}_{s}$ and its fluctuations $\chi^{2}_{s}$.
These quantities are plotted  as a function of temperature for L=32 and 60
in Fig.7(a) and 7(b), respectively. ${\cal M}^{2}_{s}$  goes to
unity at low temperatures and  it decays sharply to zero close to the
critical region.  Note that  $\chi^{2}_{s}$ displays  an
asymmetric behavior close to $T_{Z_{2}}(\approx 0.42$ for $L=60$),
where it has a sharp maximum. This indicates that the critical
exponents for $\chi^{2}_{s}$ above and below  $T_{Z_{2}}$ should be
different.  For $T>T_{Z_2}$ we fitted the MC data to
$$M_{s}^{2}\sim (\epsilon_{Z_2})^{2\beta}, \> \> \> \hbox{and} \> \> \> \>\>
\chi_{s}^{2}\sim (\epsilon_{Z_2})^{-2\gamma}\eqno(29)$$
 while for $T\leq T_{Z_2}\>\>$ $\chi_{s}^{2}$ was fitted to
$$\chi_{s}^{2}\sim (-\epsilon_{Z_2})^{-2\gamma '}.\eqno(30)$$
We extracted the critical exponents $2\beta$, $2\gamma$ and
$2\gamma\>\prime$ by a straight line fits to
$\ln({\cal M}^{2}_{s})$ vs $\ln (\epsilon_{Z_{2}}(L))$ and
$\ln \left( \chi^{2}_{s} \right)$ vs
$\ln(\left| \epsilon_{Z_{2}}(L) \right|)$
for temperatures within 10\% from the estimated $T_{Z_{2}}(L)$.
Here we used the
notation $\epsilon_{Z_{2}}=(T-T_{Z_{2}})/{T_{Z_{2}}}(L)$, with
$T_{Z_{2}}(L)$ the temperature
at which ${\cal M}^{2}_{s}$ goes steeply to zero and $\chi^{2}_{s}$
shows a maximum for a given L. The exponents obtained for the largest
lattice were
$$2\beta=0.1936(35), \> \>\>\> \>2\gamma\>\prime=1.025(79),  \> \>\>\>
\hbox{and} \>\>\> \>\>\> 2\gamma=1.82(13)\eqno(31).$$
These exponents clearly differ from the corresponding 2-D Ising
model exponents $2\beta=1/4$, $2\gamma =2\gamma \> \prime=7/2$. We note that
our chiral order parameter exponent
does agree with the value $2\beta=0.20(2)$ obtained
from MC  transfer-matrix studies of the FFXYM$^{25}$.
The results for $L=16,\, 32$ and 60 are
given in Table 4. Notice the consistency in the behavior of
${\cal M}^{2}_{s}$ which falls off to zero at about the same temperature
where $\chi^{2}_{s}$  has a maximum, indicating that the
$Z_{2}$ phase  transition takes place at $T_{Z_2}\approx$0.42.
\par
\vskip .5truecm
\section{\bf Correlation functions}
\vskip 0.3truecm
\subsection{\bf (a) U(1) correlations}
\vskip 0.3truecm
In this subsection we discuss our MC results for the
gauge invariant phase correlation functions obtained from simulations
in lattices from L=16 up to L=240, with periodic
boundary conditions. Some of these results have already been discussed in Ref.
22 and thus we will make reference to them here.
To reduce finite size effects  the lattice sizes at
each temperature were chosen such that ${L/{\xi}} \> \geq 5$. As we mentioned
in Section 3.2 this criterion has  proven to work well in the
numerical calculations of correlation functions in the
XYM. We showed in Ref. 22 that the zero-momentum
phase correlation function $g_{U(1)}(r)$ has an oscillatory behavior
with period 1/2, which comes from the Aharanov-Bohm phase factors discussed
in Section 2.3.$^{3}$ At higher temperatures we found that
the oscillatory behavior disappears, as one would expect.
As the critical temperature is approached from above the
oscillations increase in amplitude and saturate below $T_{U(1)}$.
This oscillatory behavior led us to separate the correlation
functions into two components; one for the odd and one for the even
lattice sites, with
their corresponding coherence lengths $\xi_{o}$ and $\xi_{e}$.
The MC data for the zero momentum correlation functions
was fitted to the periodic version of the {\it ansatz} given in
Eq. (16) for $T \geq T_{U(1)}$.
This procedure incorporates the periodic boundary conditions
due to the finiteness of the lattice (Eq.(22)).
We carried out unconstrained non-linear 3-parameter
fits to the data to obtain $\xi(T)$, $ \eta(T)$ and the coefficient A
for the odd and even correlation functions.  We followed our XYM approach and
fitted the MC data to the linear functions
$$\ln (g(r)) =\ln A +\ln \biggr [r^{-\eta}e^{-r/{\xi}}+(L-r)^{-\eta}
e^{-(L-r)/{\xi}}\biggr],\eqno(32)$$
varying $\eta$ until a minium for $\chi ^2$ was reached.
In Table I of Ref. 22 we gave the results
for $\xi_{o}$ and $\xi_{e}$ as a function of temperature and lattice size,
as well as the statistics of the runs.
As one gets closer to the critical temperature from above, the coherence length
increases exponentially and one needs longer simulations and larger lattices
in order to get statistically reliable data.
Furthermore, the fitting
parameter  $\eta(T)$ defined in Eq. (16) increases and oscillates rapidly,
for both the odd and the even lattices  so that
an estimate of $\eta\left(T_{U(1)}\right)$ was not attempted. The same
situation was encountered in the XYM as described in Section 3.2.
We also found that as the temperature decreases, $A_{o}$ decreases while
$A_{e}$ increases, both slowly. Far from the critical region
we got reliable results for $\xi_{o}$ and $\xi_{e}$ using lattices of size
$L\leq 60$. However, to obtain meaningful results as we got closer
to the critical region, $L$ was increased keeping the ratio
${L\over \xi} \geq 5$. For instance, we had to increase the size up to L=240
for temperatures that were about 3\% and 2\% away from T$_{U(1)}$.
In contrast, in our XYM calculations the approach to $T_{BKT}$ was only on
the order of 10\%. In Fig. 2 of Ref. 22  we showed the results for
$\xi_{o}$. Similar results are obtained for  $xi_{e}$.
We found that for the temperatures considered here $\xi_{o}<\xi_{e}$.
In the determination of the critical exponents and the critical temperature
it is crucial how one fits the data, as discussed in Ref. 46.
We first tried  a 4-parameter unconstrained non-linear fit to the MC data
of the BKT type (Eq. 17) obtaining the results,
$$\nu_{e}=0.3133(57), \>\>\> \>\> \hbox{and}
\>\>\>\nu_{o}=0.3005(6),\eqno(33)$$
which are close to 1/3. We then fixed the  values
$\nu_{e}=\nu_{o}=1/3$, and carried out a 3-parameter fit to the data
for both lattices. The quality of the fits  improved and hence
we could surmise that the correct value of this exponent may indeed be 1/3.
The first column of Table II in Ref. 22 listed the results obtained from
these fits together with their corresponding $\chi^{2}/dof$.
For completeness we also carried out fits assuming $\nu=1/2$, the standard
BKT value, and although the $\chi^{2}$ function was smaller
than when $\nu=1/3$,  we found the differences too small
to decide with absolute confidence from our data which
exponent is the correct one.   Nonetheless, as it will be seen
below, we have other arguments, for example the finite size scaling analysis
of the data, that yields better results when $\nu=1/3$,
suggesting that this may very well be the correct value.
\par
It should be stressed that doing non-linear
fits is a non-trivial matter since there is no guarantee that the
values of the estimated parameters correspond to the absolute
minimum of the $\chi^{2}$ function.
Therefore, one needs to check the results very carefully and often times
resort to different fitting procedures to cross check the results.
For instance, a good test to
check the stability of the results is to reduce the dimension of the
parameter-space by fixing one of the parameters and carrying out the fitting
procedure for the others. This process should be repeated$^{43-46}$
for a set of values of the parameter held fixed within an interval
about the value which is supposed to yield the  minimum to $\chi^{2}$.
In some instances it is worthwhile to use the value of more than
one parameter, say two, to reduce the problem to a linear fit.
For example, in the calculation of $\nu$
and $T_{U(1)}$ we carried out linear fits to
$$\ln (\xi)= \ln (A)+B(T-T_{U(1)})^{-\nu},\eqno(34)$$
with $\nu=1/2$ or $1/3$ while varying  $T_{U(1)}$ about the  value
$T_{U(1)}=0.446$, obtained from the nonlinear fits. We found that
sometimes these fits led to different values of the parameters A and B
and also to different values for the minimum of the $\chi^{2}$ function.
However, the different values of $T_{U(1)}$ extracted from these fits were not
significantly different. This uncertainty in the analysis must be
due to the complicated topology of the parameter-space and,
although our calculations are very extensive, the number of points used in the
fits with their corresponding statistical significance may not be sufficient
to obtain a clear minimum for the $\chi^{2}$ function.

Another possible source of problems relates to the size of
the errors that weight each value of $\xi(T)$ in the fits.
In our calculations  most of the errors
were on the order of $10^{-3}$ (see Table I of Ref. 22) and, because
of the small number (12) of available data points we obtained relatively
large values
for the $\chi^{2}$ function. This in turn led to small Q values, (Q being
the {\it goodness of fit}). In most cases we found Q$\leq 0.10$,
suggesting that the data did not fit the model well.
To sort out this problem
we followed standard practice by setting $\sigma^{2}_{i}=1$ and carried
out the fits again$^{51}$. We found that by doing this the
values of Q became larger than 90\% in most  cases, and the
values of  the fitted parameters remained basically the same but with
slightly smaller errors.  The
values shown in Table II of Ref. 22 were in fact obtained following this
type of analysis. One will need to have  more points to improve the quality of
the fits. To do this one needs to carry out calculations even closer
to the critical point. One of the major problems in undertaking such a
program is the lack of an efficient algorithm that could reduce effectively
the critical slowing down.
For the known algorithms that reduce  the critical slowing down
there has been the need to know in some detail
their elementary excitations. In a sense this is equivalent
to having to know the main features of the solution to the model
before an appropriate algorithm can be tailored.
\par
As an additional test of the reliability of the results for
$T_{U(1)}$ and $\nu$, we carried out a finite size scaling analysis
of the data for $\xi_{o}$ and $\xi_{e}$. For a finite system, assuming
periodic boundary conditions, the usual $T>T_{U(1)}$ finite size
scaling {\it ansatz} for a BKT transition is
$$\xi(T,L)\sim L \>F_{\xi}\Biggl( L^{-1} \exp \Bigl(B_{\xi}\epsilon^{-\nu}
\Bigr)  \Biggr),\eqno(35)$$
with $\>F_{\xi}$ the scaling function, not known {\it \` a priori}, which
must satisfy the conditions
$$F_{\xi}(x)=0, \>\>\>\>\>\hbox{as}\>\>\>\>\>x \rightarrow 0,$$
$$F_{\xi}(x)<\infty, \>\>\>\>\>\hbox{as}\>\>\>\>\>x \rightarrow \infty.$$
The idea is to find the set of parameters $B$, $\nu$ and $T_{U(1)}$
for which the data for different temperatures and lattice sizes
fall onto one curve. Fixing $\nu=1/3$, we varied the values
of $B$ and $T_{U(1)}$ about their values obtained in the previous fits.
We found that as we moved away from those values in the increasing or
decreasing directions, the data became more scattered.
However, very close to the values found
from the previous fits the data fell very close to a unique curve.
The values for which the data
collapsed approximately onto the universal curve were
$$B_{e}=1.045,
\> \> \> \> B_{o}=0.999, \> \> \>\> \>  \hbox{and}
\> \> \>\> \> T^{e}_{U(1)}=0.440,  \> \> \> \>  T^{o}_{U(1)}=0.442.\eqno(36)$$
These numbers are in rather good
agreement with the values found in the previous fits. In the inset of
Fig. 2 of Ref. 22 we
showed the results of such analyses for the odd lattice. Similar results are
obtained from the analysis of even lattices.
We see that close to the critical region the points corresponding to
lattices $L \leq 60$  are far from the universal curve.
The equivalent finite size scaling analysis fixing  $\nu=1/2$
always led to a rapidly increasing curve suggesting that closer to the
critical point it would diverge.  This analysis provides further support
in favor of $\nu=1/3$.

\par
As in the XYM analysis, we also tested  a power law fit to the
$\xi(T)$ data, and the results are given in the third column of Table II
of Ref. 22.
We find that the BKT and power law fits  are of comparable quality,
as in the XYM case. Hence, one cannot be absolutely sure from this analysis
alone which one of the two fits is the correct one. In trying to resolve
this ambiguity we also calculated $g_{U(1)}(r)$ below $T_{U(1)}$,
mostly for L=60. In  fitting the corresponding data to an algebraic
form we followed  a procedure that parallels the one discussed in Section 3.2.
The results of the analyses are presented in Tables 5A, 5B.
In Figure 8 we show the exponents $\eta_{o}(T)$ ($\square$) and
$\eta_{e}(T)$ ($\circ$) obtained from the algebraic fits to
$g_{U(1)}(r)$. A careful look at the numbers indicates that the
trend in $\eta(T)$ for both lattices is qualitatively similar to
the one found in the XYM, but they are quantitatively different.
The exponential fits to $g_{U(1)}(r)$ appear to yield better results
with $\alpha\leq 10^{-2}$, and
larger values for  $\eta(T)$ than the ones obtained
with an algebraic fit, see the results in Tables 6A and 6B. Nonetheless,
the values of $\alpha$ decreased as the lattice
sizes increased, suggesting that at the asymptotic limit the leading
contribution will mostly come from the algebraic part of the correlations.
We also calculated $\eta_{e}$ and $\eta_{o}$ at the average critical
temperature $T_{U(1)}\equiv{1\over 2}( T^{o}_{U(1)} + T^{e}_{U(1)})$
obtained from the high temperature
analyses for lattices with $L=32,\, 40,\, 48,\, 60,\, 72,\, 84$ and $96$.
The results from the finite size analysis of the algebraic fits
are summarized in Tables 5C and 5D, whereas we list the corresponding
results for the exponential fits in Tables 6C and 6D.
The resulting values for $\eta_{e}$ and $\eta_{o}$ as a function of L are
shown in the inset of Figure 8. For comparison we also show $\eta_{0}$.
Observe that $\eta_{o}$ is systematically above $\eta_{e}$ and that
the behavior of these exponents as a function of L
is qualitatively similar to those found in the XYM,
e.g. the $\eta$'s increased monotonically
with L  without appearing to saturate for the values considered.
However, the $\eta$'s do seem to reach a more asymptotic value for the FFXYM
than for the XYM.
{}From the above analysis we extracted the results,
$$\eta_{o}(T_{U(1)})=0.1955(3) \>\>\>\>\> \hbox{and}\>\>\>\>\>
\eta_{e}(T_{U(1)})=0.1875(3).\eqno(37)$$
On the other hand we get
$$\eta_{o}(T_{(U(1))})=0.2521(3)\>\>\>\>\> \hbox{and}\>\>\>\>\>
 \eta_{e}(T_{(U(1)})=0.2480(3),\eqno(38)$$
assuming the exponential fits to the correlations.
We note that in the algebraic fits the values of $\eta_{o}$ and $\eta_{e}$
are smaller than in the exponential case, as in
the XYM analysis, and clearly $\eta\neq \eta_{0}$.

Again as in Section 2.3, we carried out a check of
the universal jump relationship as applied to the  average values of the
even and odd lattice results. We found that the universal jump relation
is indeed satisfied for our FFXYM results with the jump at $T_{U(1)}$ of
$$\Upsilon (T_{U(1)})=0.37(1),\eqno(39)$$
clearly different than the XYM universal jump. Apart from giving
a strong consistency check of the set of results
obtained by independent calculations
this is a surprising finding, for there is no a priori reason why
the universality results should be valid in the FFXYM, in particular
in light of the non Ising and XY results from our study.
In the XYM the universality of the jump in $\Upsilon$ is
a consequence of an underlying universal RG result$^{36-38}$. We can then
just surmise that there may be an underlying RG argument that
will lead to an understanding of the physical properties of the FFXYM.

\vskip .5truecm
\subsection{\bf (b) $Z_2$ correlation functions}
\vskip 0.3truecm
Let us now turn to the discussion of the correlation functions for the
chiral degrees of freedom. Our study  here will be less detailed
than in the $U(1)$ case, mainly concentrating on the temperature region above
$T_{Z_{2}}$, although a few results for $T\leq T_{Z_2}$ will also be discussed.
Prior information about the chiral critical exponents is available so
that we can compare our results to them. The
calculation of the zero momentum chiral correlation functions
defined in Eq. (15) is less demanding than in the $U(1)$ case since
one expects that $\xi _{\chi}$ diverges algebraically.
The analysis of $g_{\chi}(r)$ followed a similar logic to that of the U(1)
study. The Fig. 9 inset shows $g_{\chi}(r)$ vs $r$
above and below $T_{Z_{2}}$.
The results for the coherence length $\xi_{\chi}(T)$
for different lattice sizes are given in Table 7. Figure 2 of Ref. 22
showed the
results of a power law fit of the data to
$$\xi_{\chi}\sim (\epsilon_{Z_{2}})^{-\nu_{\chi}}.\eqno(40)$$
We also found that inclusion of the errors of  $g_{\chi}(r)$ in the fits
yielded results with rather low confidence levels ($Q < 0.10$),
as in the analysis of $g_{U(1)}(r)$.
Again we followed standard procedure by assigning the same weight to each
data point, with the resulting Q values
in most cases  above 0.90, while the fitted parameters remained essentially
the same. In the $\xi_{\chi}$ case the errors in the
fitting parameters became
smaller after normalization of the errors in the $g_{\chi}(r)$ data points.
It is difficult, however,  to be absolutely sure  that
the values of the parameters  found correspond to the absolute
minimum of the $\chi^{2}$ function, as happened in the  U(1) case.
Therefore, in addition to the nonlinear
3-parameter fits, we also fitted the data to the linear function
$$\ln(\xi_{\chi})=\ln(A_{\chi})-\nu_{\chi}\ln(T-T_{Z_{2}}).\eqno(41)$$
Using a least square fit and varying the $T_{Z_{2}}$ values about 0.42
we found
$$A_{\chi}=0.33(2) \>\>\>\>\hbox{and}
\>\>\>\> \nu_{\chi}=0.80(1), \>\>\>\>\hbox{for} \>\>\>\> T_{Z_{2}}=0.430,
\eqno(42)$$
with $\chi^{2}=3.99\times 10^{-3}$.
We also performed least square fits of the data to
the straight line
$$\xi_{\chi}^{-1/\nu_{\chi} }=\tilde A_{\chi}\>T\>+\>b_{\chi},\eqno(43)$$
for fixed values of $\nu_{\chi}$
with $\tilde A_{\chi}=A_{\chi}^{-1/\nu_{\chi} }$ and
$b_{\chi}=A_{\chi}^{-1/\nu_{\chi} }\>T_{Z_{2}}$.
Remarkably, the  results obtained were
$$A_{\chi}=0.36(3), \>\>\>\> \hbox{and} \>\>\>\>T_{Z_{2}}=0.432(9),
\>\>\>\> \hbox{for} \>\>\>\> \nu_{\chi}=0.760,\eqno(44)$$
with $\chi^{2}=9.7\times 10^{-2}$.
Note that the results in Eq(44)
are close to those in Eq(42), with essentially the same $T_{Z_2}$.
In Table 8 we give
the values of the parameters extracted from the nonlinear fit.
Our result for $\nu_{\chi}$ agrees quite well with
recent finite size scaling analysis$^{20}$ that gave $\nu_{\chi}=0.85(3)$,
as well as with the MC transfer matrix calculations$^{25}$.
The advantage of the finite size scaling analysis is that $\nu_{\chi}$
 was obtained from a one parameter fit without needing a precise value for
$T_{Z_{2}}$, as in our analysis.
Therefore it appears that the  $\nu_{\chi}$ and  $T_{Z_2}$ values
obtained here from the nonlinear fits may in fact be very close to the
correct ones. It is important to emphasize
that the $T_{Z_{2}}$ found here
is consistent with the temperature at which ${\cal M}^{2}_{s}$ fell
to zero, and $\chi^{2}_{s}$ displayed a sharp maximum.
\par

In summary, our numerical analysis of the chiral degrees of freedom
led to the critical exponents
$$2\beta=0.1936(35), \>\>\>\>
2\gamma\>\prime=1.025(79), \>\>\>\> 2\gamma=1.82(13), \>\>\>\>
\hbox{and}\>\>\>\>  \nu_{\chi}=0.875(35).\eqno(45)$$
These results strongly indicate that
the $Z_{2}$ phase transition is not an Ising-like transition as had been
suspected from previous thermodynamics studies of this model.
Note that in our calculations the difference between the
$T_{Z_{2}}$ and  $T_{U(1)}$, is about  7\%, which may not be considered
as different within the size of our estimated errors. Equivalently, one
cannot rule out the possibility that
in improved numerical simulations and closer to the critical point this
difference may disappear.
\par
\vskip 1.2truecm
\chapter{\bf Conclusions and outlook}
\vskip 0.7truecm

In this paper we have presented results from extensive MC calculations of
the FFXYM. We have explicitly analyzed the separate contributions
from the  $U(1)$ and $Z_2$ freedoms.  We have extracted the
$U(1)$ and $Z_2$ critical exponents from direct calculations of their
corresponding correlation functions and selected thermodynamic properties.
We found compelling quantitative evidence that the $U(1)$ and $Z_2$
critical exponents are clearly different from those of the usual 2-D $XY$
and Ising models. We have tested our results using several consistency checks.
Our result for the $Z_2$ correlation length exponent $\nu _{\chi}$
is essentially the same as the one obtained  from other independent
numerical calculations$^{20,25}$. There are no previous calculations of the
$U(1)$ exponents to with which to compare our results. However, a reanalysis
of the experimental data leads to an $\eta$ exponent that is clearly different
from the $XYM$ result and that agrees reasonably well with the
one found in our calculations.

Our results strongly suggest  nontrivial critical behavior in the FFXYM,
in which the U(1) and $Z_2$ freedoms
are coupled in a way so as to yield novel critical exponents.
We leave for the future the question of producing the physical
understanding of the results presented in this paper, in particular the
apparent relation between the  $U(1)$ results,
${\eta_0}^{-1}(f=0)=4$, ${\nu_0}^{-1}(f=0)=2$
and our  values, ${\eta}^{-1}(f=1/2)={\eta_0}^{-1}(f=0)+1$ and
${\nu}^{-1}(f=1/2)={\nu_0}^{-1}(f=0)+1$, together with the validity
of a universal jump for the $f=1/2$ helicity modulus.

\par
In spite of the extensive calculations and detailed analyses carried out
in this paper, improved MC simulations of this system need to be done.
More data at temperatures closer to the critical point are required
to be able to obtain more accurate estimates of the critical exponents
and critical temperatures.  As noted above, among the major limitation in this
program is the
lack of a MC algorithm that could effectively minimize the critical slowing
down. Another important limitation is that calculations of correlation
functions
near criticality require ever larger lattices which sharply increases
the computer power requirements.

\ack

We thank R. Gupta and H. van der Zant for illuminating conversations.
One of us (JVJ) thanks Professor J. E. van Himbergen for his
careful reading of this paper and for his kind hospitality
at the Institute for Theoretical Physics in the University of Utrecht,
were this paper was
completed. This work was supported in part by $NSF$ grants DMR-9211339,
INT-91-18193, the NSF
Pittsburgh supercomputing center under grant PHY88081P, the
Northeastern University Research and Development Fund, and
the Dutch Organization for Fundamental Research.
The work of GRS was also partially supported by a fellowship from CONACYT,
M\' exico, by a DGAPA-UNAM project IN102291, the NSF-CONACYT grant No.
G001-1720/001328.

\endpage
\centerline {\bf REFERENCES}
\vskip 0.25truecm
\refitem{1}
J. Villain, J. Phys. {\bf C10}, 1771 (1977); {\it ibid} {\bf C10},
4793 (1977).
\refitem {2}
 E. Fradkin, B. A. Huberman and S. Shenker, Phys.
Rev. {\bf B18}, 4789 (1978).
\refitem {3}
J. V. Jos\'e, Phys. Rev. {\bf B20}, 2167 (1979).
\refitem {4}
S. Teitel and C. Jayaprakash, Phys. Rev. {\bf B27}, 598 (1983);
{\it ibid} Phys. Rev. Lett. {\bf 51}, 1999 (1983).
\refitem {5}
W. Y. Shih and D. Stroud, Phys. Rev. {\bf B28}, 6575 (1983);
{\it ibid}, Phys. Rev. {\bf B30}, 6774 (1984).
\refitem {6}
D. H. Lee, J. D. Joannopoulos, J. W. Negele, and D. P. Landau,
Phys. Rev. Lett. {\bf 52}, 433 (1984).
{\it ibid}, Phys. Rev. {\bf B33}, 450 (1986).
\refitem {7}
S. Miyashita and H. Shiba, J. of Phys. Soc. Jpn. {\bf 53}, 1145 (1984).
\refitem {8}
M. Y. Choi and S. Doniach, Phys. Rev. {\bf B31}, 4516 (1985).
\refitem {9}
Y. M. Choi and D. Stroud, Phys. Rev. {\bf B32}, 5773 (1985);
{\it ibid}, Phys. Rev. {\bf 35}, 7109 (1987).
\refitem {10}
M. Yosefin and E. Domany, Phys. Rev. {\bf B32}, 1778 (1985).
\refitem {11}
T. C. Halsey, Phys. Rev. Lett. {\bf 55}, 1018 (1985); {\it ibid},
Phys. Rev. {\bf B31}, 5728 (1985); {\it ibid},
J. of Phys. {\bf C18}, 247 (1985).
S. E. Korshunov, J. Stat. Phys. {\bf 43}, 17 (1986).
S. K. Korshunov, {\it ibid}, {\bf 43}, 1 (1986).
\refitem {12}
E. Granato and J. M. Kosterlitz J. Phys. {\bf C19}, L59 (1986);
{\it ibid}, Phys. Rev. {\bf B33}, 4767 (1986).
\refitem {13}
J. E. van Himbergen, Phys, Rev. {\bf B33}, 7857 (1986).
\refitem {14}
B. Berge, H. T. Diep, A. Ghazali, and P. Lallemand, Phys. Rev. {\bf B34},
3177 (1986).
\refitem {15}
J. M. Thijssen and H. J. F. Knops, Phys. Rev. {\bf B37}, 7738 (1988).
\refitem {16}
H. Eikmans, J. E. van Himbergen, H. J. Knops,
and J. M. Thijssen, {\it ibid} {\bf B39}, 11 759 (1989).
\refitem {17}
 G. Grest, Phys. Rev. {\bf B39}, 9267 (1989).
\refitem {18}
J. M. Thijssen and H. J. F. Knops, Phys. Rev. {\bf B42}, 2438 (1990).
\refitem {19}
E. Granato, J. M. Kosterlitz, J. Lee and M. Nightingale, Phys. Rev. Lett.
{\bf 66}, 1090 (1991).
\refitem {20}
J. Lee, J. M. Kosterlitz and E. Granato, Phys. Rev. {\bf B44}, 4819 (1991).~
\refitem {21}
J. R. Lee and S. Teitel, Phys. Rev. {\bf B46}, 3247 (1992).
\refitem {22}
G. Ram\'\i rez-Santiago and J. V. Jos\'e, Phys. Rev. Lett. {\bf 68},
1224 (1992).
\refitem {23}
A.Vallat and H. Beck, Phys. Rev. Lett. {\bf 68}, 3096 (1992).
\refitem {24}
Jong Rim Lee, Rochester University Preprint (Nov.6, 1992).
\refitem {25}
E. Granato and M. P. Nightingale,
Phys. Rev. {\bf B49}, (September) (1993).
\refitem {26}
D. Kimhi, F. Leyvraz, and D. Ariosa, Phys. Rev. {\bf B29}, 1487 (1984).
\refitem {27}
Ch. Leeman, Ph. Lerch, G. A. Racine, and P. Martinoli, Phys. Rev. Lett.
{\bf 56} 1291 (1986).
\refitem {28}
P. Martinoli, Ph. Lerch, Ch. Leeman, and H. Beck, Proc. 18th Int. Conf. on
Low Temperature Physics, Kyoto, 1987; Japanese Journal of Applied Physics,
Vol. 26 (1987) Supplement 263.
\refitem {29}
 B. J. Van Wees, H. S. J. van der Zant, and J. E. Mooij, Phys.
Rev. {\bf B35}, 7291 (1987).
\refitem {30}
H.S.J. van der Zant, H.A. Rijken, and J.E. Mooij,
Jpn. J. of Phys. {\bf 26}, Suppl. 263, 1994 (1987).
\refitem {31}
 J. P. Carini, Phys. Rev. {\bf B38}, 63 (1988).
\refitem {32}
H.S.J. van der Zant, H.A. Rijken, and J.E. Mooij,
Jour. Low Temp. Phys. {\bf 79}, 289 (1990).
\refitem {33}
B. Pannetier, J. Chaussy, R. Rammal, J. Phys. (Paris) {\bf 44},
L1853 (1983).
\refitem {34}
B. Pannetier, J. Chaussy, R. Rammal and C.
Villagier, Phys. Rev. Lett. {\bf 53}, 1845 (1984).
\refitem {35}
V. L. Berezinskii, Zh. Teor. Fiz. {\bf 61}, 1144 (1971). [Soviet Physics
JETP {\bf 32}, 493 (1971)].
\refitem {36}
J. M. Kosterlitz and D. J. Thouless, J. Phys. {\bf C6}, 1181 (1973);
J. M. Kosterlitz, J. Phys. {\bf C7}, 1046 (1974).
\refitem {37}
J. V. Jos\'e, L. P. Kadanoff, S. Kirkpatrick, and D. R. Nelson, Phys.
Rev. {\bf B16}, 1217 (1977).
\refitem {38}
D. R. Nelson and J. M. Kosterlitz, Phys. Rev. Lett. {\bf 39}, 1201 (1977).
\refitem {39}
P.  Minnhagen, Phys. Rev. Lett. {\bf 54}, 2351 (1985);
{\it ibid}, Phys. Rev. {\bf B32}, 7548 (1985).
\refitem {40}
A. F. Hebard and A. T. Fiory, Phys. Rev. Lett. {\bf 50}, 1603 (1983);
A. T. Fiory and A. F. Hebard, Phys. Rev. {\bf B28}, 5075 (1983).
\refitem {41}
M. Tinkham, D. W. Abraham, and C. J. Lobb, Phys. Rev. {\bf B28}, 6578
(1983).
\refitem {42}
R. A. Webb, R. F. Voss, G. Grinstein, and P. M. Horn, Phys. Rev. Lett.
{\bf 51}, 690 (1983).
\refitem {43}
R. Gupta, J. DeLapp, G. Betrouni, G. C. Fox, C. F. Baillie and J.
Apostolakis, Phys. Rev. Lett. {\bf 61}, 1996 (1988).
\refitem {44}
U. Wolf, Nucl. Phys. {\bf B322}, 759 (1989).
\refitem {45}
R. G. Edwards, J. Goodman, and A. D. Sokal,  Nucl. Phys. {\bf B354},
289 (1991).
\refitem {46}
R. Gupta and C. F. Baillie, Phys. Rev. {\bf B45}, 2883 (1992).

\refitem {47} See also
J. F. Fern\'andez, M. F. Ferreira, and J. Stankiewicz,
Phys. Rev. {\bf B34}, 292 (1986).
\refitem {48}
J. V. Jos\'e, G. Ram\' \i rez-Santiago and H. S. J. van der Zant,
Physica B+C (in print).
\refitem {49}
A. L. Scheinine, Phys. Rev. {\bf B39}, 9368 (1989).
\refitem {50}
D. B. Nicolaides, J. Phys. {\bf A24}, L231 (1991).
\refitem {51}
W. Press et al., {\it Numerical Recipes}, Cambridge University Press. Cambridge
(1986).
\endpage
\title {\bf Figure captions}
Figure 1. Possible trajectories $\Gamma_{A}(\vec r, \vec r\>\prime)$ and
$\Gamma_{B}(\vec r, \vec r\>\prime)$ used in evaluating
$g_{U(1)}(\vec r, \vec r\>\prime, \Gamma_{A})$
and $g_{U(1)}(\vec r, \vec r\>\prime, \Gamma_{B})$.
\vskip 0.5truecm
Figure 2. Correlation length $\xi_0$ as a function of temperature.
The circles denote the results from the fits to the
correlation function given in Eq.(16), with their corresponding statistical
errors. The continuous line is the result of a fit of the data to the
expression for $\xi_0(T)$ given in Eq(17).
The specific values of the fitting parameters is given in Table 1A.
The dashed line was obtained using the fitted parameters from Ref. 43,
\vskip 0.5truecm
Figure 3. (a) Results for  $\eta _0(T)$ obtained from algebraic fits
to $g_0(r)$ for L=60. The dotted line is the spin-wave results. (b)
shows $\eta _{0c}(T=T_{BKT})$ versus L.
\vskip 0.5truecm
Figure 4. (a) Finite size analysis of  $\Upsilon^{o}(T=T_{BKT})=
\Upsilon^{o}_{T_{BKT}}$, with $T_{BKT}$ obtained from the $g_0(r)$ analysis,
 as a function of $1/L$ for $L=96,\> 84,\> 72, \> 60, \>48,$ and $32$.
The straight line is a linear fit to the data, with
the $L=\infty$ extrapolated value indicated. (b) The same as in (a) for
$\Upsilon(T=T_{U(1)})=\Upsilon_{T_{U(1)}}$.
\vskip 0.5truecm
Figure 5. $\Upsilon$  as a function of $T$ for
different $L$ sizes. Note that  the data for $L=32$ and $60$
almost fall on top of each other, suggesting that the $L$ dependence
is almost negligible for $L\geq 32$.
\vskip 0.5truecm
Figure 6. Staggered magnetization due to the superconducting
currents defined in Eq. (5)
as a function of $T$ for $L=32$ and $60$. The fall off to zero
occurs at about $T=0.42$.
\vskip 0.5truecm
Figure 7. Staggered chiral magnetization square $M^2_{s}$ (a) and
susceptibility $\chi^{2}_{s}$ (b)
as a function of $T$ for $L=32$ and $60$. The results for both lattice sizes
are essentially on top of each other. Note that
${\cal M}^{2}_{s}$ goes  to zero at essentially the same $T$ at
which $\chi^{2}_{s}$ has a maximum with
$T_{Z_2}\approx 0.42$. Observe the asymmetric behavior of
$\chi^{2}_{s}$ about critical temperature.
\vskip 0.5truecm
Figure 8. Results for $\eta_{o}(T)$ ($\square$) and $\eta_{e}(T)$ ($\circ$)
obtained from algebraic fits to $g_{U(1)}(r)$ for L=60.
 The inset shows the results for the finite size analysis
for the critical exponents $\eta_{0}$ ($\square$), $\eta^{e}$
($\circ$) and $\eta^{o}$ ($\times$), at criticality.
\vskip 0.5truecm
Figure 9. Data for the chiral coherence length $\xi_{\chi}(T)$ ($\times$)
calculated from $g_{\chi}(r)$. The solid line is the fit to the form
given in Eq(20). The inset shows $g_{\chi}(r)$ as a function
of $r$ for temperatures above and below $T_{Z_{2}}$.
\endpage

\title{\bf Table captions}
\smallskip
{\bf Table 1.}
Critical exponents and critical temperatures for the XYM.
(A) In the first and second columns
we give the results obtained from fitting  the $T>T_{BKT}$ $\>$ $\xi_0$ data
to Eq(17) and Eq(20), respectively. The third column gives the parameters
obtained from an algebraic fit to $g_0(r)$ at $T=T_{BKT}$.
(B) Same as in (A) for the first two columns
with results from Ref. 46, obtained from
unconstrained  4-parameter nonlinear fits. (C) The
first two lines are the
$\eta_0$ exponents obtained from a high temperature analysis
of the susceptibility, $\chi$ (from Ref.  44). The third and fourth lines
give  $\eta_0$ obtained from MC
renormalization group calculations (Ref. 46).
\smallskip
{\bf Table 2.} (A)  Results from fits to the data for the correlation
function $g_0(r)$ to the form given in Eq.(18) for
L=60. (B) $L$ dependence of $\eta_0$ and $C_0$ at $T=T_{BKT}$.
(C) Results from exponential fits to the data of $g_0(r)$  for $T<T_{BKT}$
for L=60. (D) $L$ dependent results for $\eta_0$, $\alpha_0$ and $C_0$ from
exponential fits to the correlations at $T=T_{BKT}$.
\smallskip
{\bf Table 3.}
(A) Gives the finite size results for
$\Upsilon^{o}_{BKT}$,  while (B) gives the corresponding results for
$\Upsilon_{T_{U(1)}}$.
\smallskip
{\bf Table 4.}
Magnetic critical exponents for the $Z_{2}$ transition for
$L=16,32$ and $60$. See text for definition of the parameters.
\smallskip
{\bf Table 5.}
Results from algebraic fits to $g_{U(1)}(r)$, (A) odd and (B) even
lattices for $T<T_{U(1)}$ with $L=60$. Results of a finite size analysis
of the algebraic fits to $g_{U(1)}(r)$ for the odd (C) and even (D)
lattices at $T=T_{U(1)}$.
\smallskip
{\bf Table 6.}
Results from exponential fits to $g_{U(1)}(r)$ for the odd (A) and
even (B) lattices for $T<T_{U(1)}$ with $L=60$.
Results from a finite size analysis of exponential fits to
$g_{U(1)}(r)$ for the odd (C) and even (D) lattices at $T=T_{U(1)}$.
\smallskip
{\bf Table 7.}
Results for $\xi_{\chi}(T)$ obtained from $g_{\chi}(r)$
at different temperatures and lattice sizes.
\smallskip
{\bf Table 8.}
Exponent $\nu_{\chi}$ and critical temperature $T_{Z_{2}}$
obtained from nonlinear fits to a power law divergence.

\end